\documentclass[twocolumn]{article}
\usepackage[a4paper, margin=1.8cm]{geometry}

\usepackage{graphicx} 
\usepackage{multirow}
\usepackage{natbib}
\usepackage{bm}
\usepackage{amsmath,amssymb,graphicx,mathtools,hyperref,multicol,booktabs,float,mathrsfs,textcomp,xcolor,amsfonts}

\usepackage{amsthm}

\newcommand\norm[1]{\lVert#1\rVert}
\usepackage{tikz}
\usepackage{siunitx}
\usepackage{comment}
\usepackage[scr]{rsfso}
\newtheorem{remark}{Remark}
\newtheorem{result}{Result}
\newtheorem{theorem}{Theorem}
\newtheorem{corollary}{Corollary}
\newtheorem{assumption}{Assumption}
\newcommand{\R}{\mathbb{R}}

\definecolor{orange2}{RGB}{255,165,0}
\definecolor{violet2}{rgb}{0.93, 0.51, 0.93}
\definecolor{green2}{rgb}{0.2, 0.75, 0.2}
\newcommand{\norev}[1]{ {#1} }

\newcommand{\revone}[1]{ {#1} }

\newcommand{\revtwo}[1]{ {#1} }

\newcommand{\revthree}[1]{ {#1} }

\usepackage[font=footnotesize,labelformat=simple]{subcaption}

\begin{document}

\title{Set-membership identification of continuous-time\\ MIMO systems via Tustin discretization }

\author{Vito Cerone, Sophie M. Fosson, Simone Pirrera, Diego Regruto\thanks{Corresponding author D.~Regruto, email: diego.regruto@polito.it, Tel. +39-090-011-7077.}}
\date{}

\maketitle

\begin{abstract}                          
In this paper, we deal with the identification of continuous-time systems from sampled data corrupted by unknown but bounded errors. 
A significant challenge in continuous-time identification is the estimation of the input and output data derivatives. In this paper, we propose a novel method based on set-membership techniques and Tustin discretization, which overcomes the derivative measurement problem and the presence of bounded errors affecting all the measured signals.
First, we derive the proposed method and prove that it becomes an affordable polynomial optimization problem. Then, we present some numerical results based on simulation and experimental data to explore the effectiveness of the proposed method.
\end{abstract}

\maketitle
\thispagestyle{empty}
\pagestyle{empty}



\section{Introduction} 
Identification of continuous-time (CT) linear time-invariant (LTI) systems is a long-standing problem that has recently attracted considerable renewed attention. Although discrete-time (DT) models are widely employed thanks to their compliance with digital computing, building CT models is still attractive in several aspects. Since most real-world systems are intrinsically continuous in time, CT modeling, prone to include partial a-priori knowledge of the parameters, is the correct paradigm to obtain physically interpretable model; see, e.g., \cite{unb90,gar03,gar15}. Furthermore, CT models are preferable to handle irregularly sampled data, see \cite{lar07,mu15}, and to account for arbitrary time delays, which may lead to non-minimum phase DT models; see \cite{che18,che20}.
Finally, we notice that the CT representations of the plant is the preferred modeling paradigm in the context of frequency domain robust control based on loop-shaping design, $\mathcal{H}_\infty$ optimization, and $\mu$-synthesis (see, e.g., \cite{zhou97,doyle09}), although DT counterparts of such techniques have also been derived.

Although we usually model CT systems through differential equations, such a mathematical description makes the identification problem challenging because it requires computing time derivatives from sampled and noisy data; see, e.g., \cite{you80}, \revthree{\cite{pintelon1997identification}, }and \cite{gar03} for a detailed discussion.
%
In the literature, several methods address the derivative measurement problem. A classical approach to CT single-input, single-output (SISO) systems is to apply state variable filters on the input and output data to approximate the derivatives; state variable filtering is a branch of the instrumental variable methods developed by \cite{young65a,young70}. 
\cite{joha94} employs low-pass filters to perform a model transformation that does not require estimation of the derivatives. The proposed transformation provides an equivalent linear model whose parameters can be identified by linear regression.
Beyond state variable filters and model transformation, we also mention integration methods, see \cite{sagara90,trig96}, and Fourier modulating functions, see \cite{pear94,rao98}.

A drawback of the approaches mentioned above is that they are not conceived to identify CT multiple-inputs multiple-outputs (MIMO)  systems. Instead, subspace methods are used, as discussed in \cite{ouv12}. Initially developed for DT systems by \cite{over96}, subspace identification is extended to CT MIMO systems in several works, see, e.g., \cite{hav97,mer07,mer12}.

%

All the approaches mentioned above consider either the noise-free case or a stochastic description for the measurement noise affecting the collected data. Much less attention has been paid so far to the case of CT identification in the presence of unknown but bounded (UBB) errors corrupting the input-output measurements. The use of the bounded-error (or set-membership) characterization of the measurement errors is motivated, in particular, in those cases where either  prior statistical information
is not available, or the errors are better characterized in a deterministic
way (e.g., systematic and class errors in measurement
equipment, rounding and truncation errors in digital devices).
Based on the UBB uncertainty description, an alternative paradigm called
bounded-error or set-membership (SM) identification has progressively
emerged in the last four decades. We refer interested readers to the book by \cite{mila96} for an introduction to the subject. 
%

In this paper, we consider the set-membership identification of CT LTI systems from sampled data under the assumption that the input and the output samples are corrupted by bounded noise. We refer to this problem as the set-membership errors-in-variables (SM-EIV) problems.

Although several contributions are available in the literature addressing the SM-EIV problem for DT systems, \revthree{see, e.g., \cite{1987Ano,ver92,cer93a,geng2010worst,cer12,cer18,cerone24autom},  }
to the best of our knowledge, \cite{cdc22} is the first contribution addressing the SM-EIV identification in the CT case. \norev{In that paper the authors formulate the identification problem as a polynomial optimization problem (POP) and solve it through convex relaxation techniques by elaborating on the model transformation approach for SISO systems of \cite{joha94}.} 
%
%

%

In this work, we develop a novel method for SM-EIV identification of CT systems, based on Tustin discretization. In contrast to \cite{cdc22}, this approach allows us to \norev{identify MIMO systems through an algorithm of relatively low complexity.}

We summarize the main contributions as follows. Firstly, we develop and analyze the proposed Tustin-based identification method by illustrating its advantages with respect to previous approaches. Secondly, we address the problem of dealing with the discretization error. Finally, we propose several numerical tests performed on both simulated and experimental data that support the effectiveness of the proposed method.

We organize the paper as follows. \norev{In Section \ref{sec:prob_formul}, we present the mathematical formulation of the considered problem. We develop the proposed solution in Section \ref{sec:tustin} by assuming that a bound on the discretization error is a-priori known. A data-driven approach to compute such a bound is presented in Section \ref{sec:discr_estim}.} Section \ref{sec:numerical} is devoted to numerical tests;  Section \ref{sec:concl} concludes the paper.

\norev{\textbf{Notation.}~We denote $s$ and $z$ the Laplace and $\mathcal{Z}$-domain variables, respectively, while $q$ is the forward time-shift operator. Given two vectors of compatible size, $\leq$ denotes the componentwise inequality. The symbols $A_{i,j}$ and $v_{i}$ define, respectively, the $(i,j)$ entry of matrix $A$ and the $i$-th element of vector $v$. For a vector $v$, $a = \vert v \vert$ is defined componentwise by $a_i = \vert v_i \vert$. $A_{:,i}$ denotes the $i$-th column of matrix $A$. 
}

\section{Problem formulation}
\label{sec:prob_formul}
We consider a MIMO system described through a matrix transfer function $H(s) \in \mathbb{C}^{n_y,n_u}$, where $n_u$ is the number of inputs, $n_y$ is the number of outputs. Each entry $H_{m,l}(s)$ of the matrix $H(s)$ is a SISO CT \revtwo{bounded-input bounded-output (BIBO) stable} model of known order $n_{m,l}$, described by: 
\begin{equation} 
H_{m,l}(s) = \dfrac{\sum_{j=0}^{n_{m,l} -1} \beta^{(m,l)}_j s^j}{s^{n_{m,l}} + \sum_{j=0}^{n_{m,l}-1} \alpha^{(m,l)}_j s^j} 
 \label{eq:tf_mimo} \end{equation}
where $l = 1,\dots,n_u$, $m = 1, \dots,n_y$.

The input of the system is a vector of CT \revtwo{bounded} signals $u(t) \in \mathbb{R}^{n_u}$; the output is a vector of CT signals $y(t) \in \mathbb{R}^{n_y}$. Input and output signals are sampled with constant rate $T_s$ and corrupted by the noise sequences $\eta(k)  \in \mathbb{R}^{n_y}$ and $\epsilon(k) \in \mathbb{R}^{n_u}$. The data used to perform the CT model identification are the samples
\norev{
\label{eq:eiv_mimo}
\begin{align}
    \tilde{u}(k) &= u(k) + \epsilon(k), \quad \tilde{y}(k) = y(k) + \eta(k).
    \label{eiv_mimo}
\end{align}
where, for each $k = 1,\dots,N$, 
\begin{equation} 
    u(k) \doteq u(t=kT_S), \qquad    y(k) \doteq y(t=kT_S).
    \label{samples}
\end{equation}
}
\norev{We consider noise sequences satisfying the following assumption: 
\begin{assumption}
    The noise signals are bounded according to
    \begin{subequations}\begin{align}
        &\underline{\Delta_\epsilon}(k) \leq \epsilon(k) \leq \overline{\Delta_\epsilon}(k), \\
        &\underline{\Delta_\eta}(k) \leq \eta(k) \leq \overline{\Delta_\eta}(k). \label{noise_bnds_mimo}
    \end{align}\end{subequations}
    where, for each $k=1,\dots,N$, $\underline{\Delta_\epsilon}(k),\overline{\Delta_\epsilon}(k) \in \R^{n_u}$, and $\underline{\Delta_\eta}(k), \overline{\Delta_\eta}(k) \in \R^{n_y}$ are given constants.
\end{assumption}
}
The objective of the proposed approach is to estimate the parameters of each transfer function $H_{l,m}(s)$, namely $\theta^{(l,m)} = [\alpha_0^{(l,m)}, \dots, \alpha_{n_{l,m}-1}^{(l,m)}, \beta_0^{(l,m)}, \dots, \beta_{n_{l,m}-1}^{(l,m)}] \in \mathbb{R}^{2 n_{l,m}}$. \norev{We denote $\theta \doteq [\theta^{(1,1)},$ $\dots,$ $ \theta^{(n_u,n_y)}] \in \R^{n_\theta}$ the vector of all the model's parameters, with $n_\theta = \sum_{l=1}^{n_u} \sum_{m=1}^{n_y} 2\,n_{l,m}$.}\\

According to the SM estimation theory, we define the feasible parameters set (FPS) $\mathcal{D}$ as the set of all the values of parameters $\theta$ coherent with the collected data and the prior information on the system and the noise. For the problem considered in this work, the implicit definition of the FPS is given by 
\norev{
\begin{equation}
\begin{aligned}
    \mathcal{D} &= \{ \theta \in \mathbb{R}^{n_\theta}: \\
    & y_m(t) = \sum_{l=1}^{n_u} h_{l,m}(t|\theta)*u_l(t),\quad m = 1,\dots,n_y,\\
        &\tilde{u}(k) = u(k) + \epsilon(k), \quad \underline{\Delta_\epsilon}(k) \leq \epsilon(k) \leq \overline{\Delta_\epsilon}(k) \\ 
        &\tilde{y}(k) = y(k) + \eta(k), \quad  \underline{\Delta_\eta}(k) \leq \eta(k) \leq \overline{\Delta_\eta}(k)\\
        & \text{ for }\quad k = 1,\dots,N \}
\end{aligned}
\label{eq:fps_generic}
\end{equation}
}
where $\norev{h_{l,m}(t|\theta) = \mathcal{L}^{-1}\{[H(s)]_{l,m}\}}$ is the impulse response of the system \norev{(depending on the system's parameters $\theta$)} and $*$ is the convolution operator.
In this work, we refer to the following general formulation of the SM identification problem:
\begin{equation}
    \hat{\theta}_{SM} = \arg\min_{\theta\in\mathcal{D}}\mathcal{J}(\theta)  \label{eq:sm_generic}
\end{equation}
where $\mathcal{J}(\theta)$ is a generic multivariate polynomial function. Note that such a class is sufficiently general to account for most cost functions commonly used in system identification, including, e.g., $\ell_1$-norm, $\ell_2$-norm, and $\ell_{\infty}$-norm of prediction/simulation error.  
In particular, by setting $\mathcal{J}(\theta)=\theta_i$ and $\mathcal{J}(\theta)=-\theta_i$, we evaluate the parameter uncertainty intervals (PUIs), defined as (see, e.g., \cite{cer12})
\begin{equation}
    \text{PUI}_{\theta_i} = [\underline{\theta}_i, \overline{\theta}_i ],~~~i=1,\dots,n_\theta
\end{equation}
where
\begin{subequations}
\begin{align}
    \underline{\theta}_i = \min \theta_i \quad \text{ s.t. } \, \theta \in \mathcal{D}, \label{eq:min_generic} \\
    \overline{\theta}_i = \max \theta_i \quad \text{ s.t. } \, \theta \in \mathcal{D}.   \label{eq:max_generic}
\end{align} 
\label{eq:generic_pop}
\end{subequations}
Given the PUIs, it is possible to evaluate the Chebyshev center  $\theta^C$ of $\mathcal{D}$ in the $\ell_\infty$-norm:
\begin{equation}\label{cheby}
    \theta^C = \min_{\theta' \in \mathbb{R}^{n_\theta}} \max_{\theta \in \mathcal{D}} \norm{\theta' - \theta}_\infty
\end{equation}
which is given by the \textit{central estimate} 
\begin{equation}
    \theta_j^C = \dfrac{\underline{\theta}_j + \overline{\theta}_j}{2}, \quad j = 1,\dots,n_\theta.
    \label{cheby2}
\end{equation}
\norev{The Chebyshev center in \eqref{cheby} is an optimal SM estimator in the sense that it minimizes the maximum possible distance from the true parameter value.}
We refer the reader to \cite{celapire14} and references therein for a detailed discussion on central and conditional central estimation in the SM EIV framework.

\norev{
\section{Identification via polynomial optimization}
\label{sec:tustin}
}
In this section, we develop a novel CT SM identification method based on Tustin discretization and polynomial optimization. 

\subsection{Tustin discretization approach}

\norev{A possible approach to this problem is the indirect identification strategy, in which the relationship between the data samples is described by a DT transfer function $T(z)$. By assuming a known mapping between $T(z)$ and $H(s)$, the CT model can be recovered; see, e.g., \cite{unb90} for details.
However, as noted by \cite{gar15}, this approach suffers from some drawbacks: the mapping from DT to CT may be ill-conditioned, and the relative degree of the obtained CT models may be physically inconsistent.}
\revone{Moreover, in the SM setting, if we estimate the parameters via DT SM identification, as in \cite{cer12}, and then map them to CT via discrete-to-continuous transformation, we introduce conservativeness because the PUIs of the DT model are larger than the true ones.}
\norev{For these motivations,} this work proposes a novel approach based on a one-step strategy that concurrently builds the CT model and a DT approximation by solving a suitable POP. 


\norev{First of all, we derive the DT model representation by applying the Tustin bilinear transformation 
\begin{equation}\label{eq:replace}
    s = \dfrac{2}{T_s} \dfrac{z-1}{z+1}.
\end{equation} 
to the CT model; see, e.g., \cite{astr97}. $T_s$ represents the discretization period.}

\revthree{Given a CT model $H(s)$, an input signal $u(t) \in \R^{n_u}$, and a discretization period $T_s$, we quantify the error introduced by the DT Tustin approximation $T(z)$ of $H(s)$ by defining the discretization error as
\begin{equation}
    \delta(k) \doteq  y(k) - \tau(k)
\end{equation}
where $\tau(k)$ is the output of $T(z)$ with input $u(k)=u(k\, T_s)$, and $y(k) = y(k\, T_s)$ is the sampled output of $H(s)$ when the input is $u(t)$.
}

\begin{result}\label{result1}
\norev{For a SISO model 
\begin{equation}\label{eq:sisotf}
    H(s) = \dfrac{\sum_{j=0}^{n -1} \beta_j s^j}{s^{n} + \sum_{j=0}^{n-1} \alpha_j s^j},
\end{equation}
the noise-free output samples ${y}(k)$ are given by
\begin{equation}\label{eta_delta_additive_1}
{y}(k) = \tau(k) +  \delta(k)
\end{equation}
}
\begin{equation}
     \tau(k) = \sum\limits_{j=0}^{n}\xi_j u(k-n+j) - \sum\limits_{j=0}^{n-1}\gamma_j \tau(k-n+j)
    \label{eq:in_out_dt_1}
\end{equation}
\norev{where, for $j= 0,\dots,n$ and $i = 0,\dots,n-1$:}
\begin{equation} \label{eq:dt_from_ct_frac_1}
     \xi_j = \frac{b_j(\beta)}{a_n(\alpha)}, \qquad 
    \gamma_i = \frac{a_i(\alpha)}{a_n(\alpha)}
\end{equation}
\begin{subequations}\label{eq:def_ab_fun}
\begin{align}
    b_j(\beta) &= \sum\limits_{i=0}^{n-1} \beta_i 2^i T_s^{n-i} c_{j}^{(i)}\\
    \label{b_affexpr}
    a_j(\alpha) &= 2^n c_{j}^{(n)} + \sum\limits_{i=0}^{n-1} \alpha_i 2^i T_s^{n-i} c_{j}^{(i)}\\
    \label{a_affexpr}
    c^{(i)}_j &= \sum_{k=0}^j \binom{i}{k}(-1)^{i-k}\binom{n-i}{j-k}.
\end{align}
\end{subequations}
\end{result}

\begin{proof}
\revthree{
By replacing \eqref{eq:replace} in \eqref{eq:sisotf}, and using the definition of the functions $a_j(\alpha)$ and $b_j(\beta)$, we get
\begin{equation} \begin{aligned}
    T(z)&  = \dfrac{\sum\limits_{i=0}^{n-1} \beta_i \left(\dfrac{2}{T_s} \dfrac{z-1}{z+1}\right)^i }{\left(\dfrac{2}{T_s} \dfrac{z-1}{z+1}\right)^n + \sum\limits_{i=0}^{n-1} \alpha_i \left(\dfrac{2}{T_s} \dfrac{z-1}{z+1}\right)^i } =\\
    & = \dfrac{\sum\limits_{i=0}^{n-1} \beta_i 2^i (z-1)^i T_s^{n-i} (z+1)^{n-i}}{2^n(z-1)^{n} + \sum\limits_{i=0}^{n-1}\alpha_i 2^i (z-1)^i T_s^{n-i}(z+1)^{n-i}} \\
    &= \left({\sum\limits_{j=0}^{n} b_j(\beta) z^j }\right) \left({\sum\limits_{j=0}^{n} a_j(\alpha) z^j}\right)^{-1}.
    \label{eq:tf_dt_ab}
\end{aligned}\end{equation}
If we properly select the order of the system, then $a_n(\alpha) \neq 0$ and we can normalize \eqref{eq:tf_dt_ab} with respect to $a_n(\alpha)$. The obtained expression is a standard DT LTI model of order $n$ whose parameters are given in \eqref{eq:dt_from_ct_frac_1} and whose input-output behavior is \eqref{eq:in_out_dt_1}.
}
\end{proof}

\revthree{
\begin{remark}
    Result \ref{result1} provides an exact mathematical description of the mapping between the input and output samples, expressed as a function of the CT model parameters $\theta$ and the discretization error $\delta$.
    \\
\end{remark}
}

\revone{
\begin{assumption}\label{ass:discr_err}
    The discretization error is bounded as
    \begin{equation}
        \left|\delta(k)\right| \leq \Delta_\delta(k)
     \end{equation} 
     where, for all $k=1,\dots,N$, $\Delta_\delta(k)$ is given. Moreover, the bounds $\Delta_\delta(k)$ are uniquely determined by
     \begin{equation}\label{tustin_bound} 
         \Delta_\delta(k) = d\, \phi_k(\tilde{y}(1),\dots,\tilde{y}(N),\tilde{u}(1),\dots,\tilde{u}(N)),
     \end{equation}
     where $\phi_k$ are real-valued functions and $d$ is a real constant. {Equation \eqref{tustin_bound} allow us to cover different interesting cases}, e.g., (i) the \textit{uniform bound} $\Delta_\delta(k) = d$ for all $k = 1,\dots,N$, (ii) the \textit{relative output (input) bound} $\Delta_\delta(k) = d \,\vert \tilde{y}(k) \vert$ ($\Delta_\delta(k)= d \,\vert \tilde{u}(k) \vert$) for all $k = 1,\dots,N$, and (iii) the \textit{relative differential output (input)  bound} $\Delta_\delta(k) = d\, \vert \tilde{y}(k) - \tilde{y}(k-1) \vert$ ($\Delta_\delta(k)= d \,\vert \tilde{u}(k) - \tilde{u}(k-1) \vert$) for all $k = 1,\dots,N$. We discuss the problem of computing $d$ from available data in Section \ref{sec:discr_estim}.
     \\
\end{assumption}
} 

\revtwo{
\begin{result}
    Let $H(s)$ be BIBO stable and $u(t)$ bounded. Then, the discretization error is bounded. Therefore, for each $k=1,\dots,N$, there exists a constant  $\Delta_\delta(k)$ such that $\vert \delta(k) \vert \leq \Delta_\delta(k)$.
\end{result}
\begin{proof}
    $\delta(k)$ is bounded if both $y(k)$ and $\tau(k)$ are bounded. $y(k)$ is bounded because it is obtained by sampling the output of $H(s)$, which is BIBO stable, when a bounded input $u(t)$ is applied. Similarly, $\tau(k)$ is bounded because it is the output of $T(z)$ excited by the bounded input sequence $u(k)$, and $T(z)$ is BIBO stable because the Tustin transform preserves stability regardless of the sampling rate.  
\end{proof}
}

\norev{
\subsection{Identification of MIMO systems}
We start by describing the relation between input-output samples using a matrix of DT transfer functions 
\begin{equation}
   T_{m,l}(z) = 
   \dfrac{\sum\limits_{j=0}^{n_{m,l}} \xi_j^{(m,l)} z^j }{z^{n_{m,l}} + \sum\limits_{j=0}^{n_{m,l}-1} \gamma_j^{(m,l)} z^j}, \quad 
    \label{eq:tf_miso_dt}
\end{equation}
for $m = 1,\dots,n_y$, $l=1,\dots,n_u$,} obtained by Tustin discretization of the corresponding CT models $H_{m,l}(s)$ in equation~\eqref{eq:tf_mimo}. Therefore,  equations \eqref{eq:dt_from_ct_frac_1} hold for each of the \norev{$n_u\times n_y$} transfer functions to identify, that is 
\begin{equation}
    \left.\begin{aligned}
    & \xi_j^{(m,l)} [a_n(\alpha^{(m,l)})] = b_j(\beta^{(m,l)}) \\
    & \gamma_j^{(m,l)} [a_n(\alpha^{(m,l)})] = a_j(\alpha^{(m,l)})  
     \end{aligned} \right\} ~ \begin{aligned} \norev{m }&\norev{= 1,\dots,n_y,} \\ l&=1,\dots,n_u, \\ j &= 0,\dots,n_{m,l}-1. \end{aligned}
     \label{eq:dt_from_ct_params_poly_miso}
\end{equation}
By using the transfer function \norev{$T_{m,l}(z)$} to describe the CT system \norev{$H_{m,l}(s)$}, we implicitly introduce a discretization error \norev{$\delta_{m,l}(k)$.} 
According to \cite{cer18}, \norev{for each $m = 1,\dots,m$ and $l=1,\dots,n_u$}, we denote the outputs of the SISO DT transfer functions \norev{$T_{m,l}(z)$}, called \textit{partial outputs}, with the symbol \norev{$z(k) \in \R^{n_y,n_u}$}. Then, from \eqref{eq:tf_miso_dt}, we obtain the following relations:
\allowdisplaybreaks
\begin{subequations} \begin{align}
\begin{split}
    & z_{m,l}(k) + \sum_{i=1}^{n_{m,l}} \gamma_i^{(m,l)} z_{m,l}(k+i-n_{m,l}) = \\ &\quad =\sum_{i=1}^{n_{m,l}} \xi_i^{(m,l)} u_l(k+i-n_{m,l}), \quad l=1,\dots,n_u,
\end{split}\\
   & \tau_m(k) \doteq \sum_{l=1}^{n_u} z_{m,l}(k).
\end{align} \end{subequations}
for $m=1,\dots,n_y$, $k = 1,\dots,N$. Each element $y_m(t)$ of the noise-free sampled output $y(k)$ of the MIMO CT system is given by:
\begin{equation}
    y_m(k) = \sum_{l=1}^{n_u} z_{m,l}(k) + \delta_{m,l}(k) = \tau_m(k) + \delta_m(k),
    \label{eq:miso_discr_err}
\end{equation}
where 
\begin{equation}
    \delta_m(k) \doteq \sum_{l=1}^{n_u} \delta_{m,l}(k)
    \label{def_delta_miso}
\end{equation}
accounts for the overall effect of the discretization error.
Finally, considering the additive noise $\eta(k)$ corrupting the output measurement, we have
\begin{equation}\label{eta_delta_additive_miso}
\tilde{y}(k) = \tau(k) + \eta(k) + \delta(k).
\end{equation}
where $\tau(k) = [\tau_1(k),\dots,\tau_{n_y}(k)]$. Therefore, \revone{under Assumption \ref{ass:discr_err}}, we obtain
\begin{equation}
    \underline{\Delta_\eta}(k)-\revone{\Delta_\delta(k)} \leq \tilde{y}(k) -  \tau(k) \leq \overline{\Delta_\eta}(k) +\revone{\Delta_\delta(k)}.
    \label{eq:output_bnds_miso}
\end{equation}

For what concerns the inputs, based on equations \eqref{eiv_mimo} and \eqref{noise_bnds_mimo}, we obtain the following constraints
\begin{equation}
    \underline{\Delta_\epsilon}(k) \leq \tilde{u}(k) - u(k) \leq \overline{\Delta_\epsilon}(k)
\end{equation}
for $k = 1,\dots,N$. We can now write the optimization problem to be solved to perform SM identification of the CT MISO system:
\allowdisplaybreaks 
\begin{subequations}     \label{min_tustin_miso} \begin{align}
    & \min_{\alpha,\beta,\gamma,\xi,u,z} \mathcal{J}(\alpha,\beta,\gamma,\xi,u,z)  \\
    & \quad \text{\rm subject to:} \notag \\
    & \quad z_{m,l}(k) + \sum_{i=1}^{n_{m,l}} \gamma_i^{(m,l)} z_{m,l}(k+i-n_{m,l}) = \label{eq:csn_io} \\ &\quad =\sum_{i=1}^{n_{m,l}} \xi_i^{(m,l)} u_l(k+i-n_{m,l}), \quad m=1,\dots,n_y, \notag \\
    & \qquad l=1,\dots,n_u,\quad k=n_{m,l}+1,\dots,N \notag \\
    & \quad \left.\begin{aligned}
    & \xi_j^{(m,l)} [a_n(\alpha^{(m,l)})] = b_j(\beta^{(m,l)}) \\
    & \gamma_j^{(m,l)} [a_n(\alpha^{(m,l)})] = a_j(\alpha^{(m,l)})  
     \end{aligned} \right\} ~ \begin{aligned} {m }&{= 1,\dots,n_y,} \\ l&=1,\dots,n_u, \\ j &= 0,\dots,n_{m,l}-1. \end{aligned} \label{eq:cns_param}\\
    & \quad \underline{\Delta_\eta}(k)-\revone{\Delta_\delta(k)} \leq \tilde{y}(k) - \sum_{l=1}^{n_u}z_{:,l}(k) \leq \overline{\Delta_\eta}(k) +\revone{\Delta_\delta(k)} \notag \\
    & \quad \underline{\Delta_\epsilon}(k) \leq \tilde{u}(k) - u(k) \leq \overline{\Delta_\epsilon}(k) \label{eq:cns_bnd} \\
    & \qquad k = 1,\dots,N \notag
\end{align}\end{subequations}
\revone{
\begin{remark}\label{rmk:subset_F_D}
    Note that Problem \eqref{min_tustin_miso} differ from \eqref{eq:sm_generic}. The main difference arises from the introduced reformulation, which required lifting the original problem into a higher-dimensional space. This was achieved by introducing auxiliary slack variables, which allowed \eqref{eq:fps_generic} to be transformed into a set of polynomial equations (and the use of polynomial optimization techniques, as we will detail in Section~\ref{sec:pop_relax}). As a result, \eqref{eq:fps_generic} is a subset of the projection of the feasible set of \eqref{min_tustin_miso} onto the original parameter space, with equality achieved if the bounds in Assumption 2 are tight.
\end{remark}
}

\revone{
\subsubsection{Exploiting additional a-priori information}
The proposed formulation naturally accommodates the inclusion of additional a-priori information by incorporating constraints into Problem~\eqref{min_tustin_miso}.

For example, the relative degree of the $(m,l)$-th transfer function can be fixed to a specific value $1 < r_{m,l} \leq n_{m,l}$ by setting the corresponding high-order numerator coefficients in $s$ to zero: $\beta^{(m,l)}_j = 0$ for $j = n_{m,l}, \dots, n_{m,l} - r_{m,l}$.
Similarly, bounds on the DC gain or individual parameters can be imposed using linear inequality constraints.\\
Additionally, the Routh-Hurwitz criterion can be employed to ensure system stability. Since the construction of the Routh array involves only addition and multiplication operations, it yields polynomial conditions on the parameters $\alpha$. Thus, including these relations does not change the nature of the problem, which remains polynomial. This can be interpreted as the CT counterpart of the results by~\cite{cerone_enforcing_2011}.

The effect of introducing such constraints is a reduction in the size of the feasible set, which in turn guarantees an improvement in the quality of the estimated parameters in terms of the PUI's size.
} 


\norev{
\subsection{Algorithm's Complexity Analysis}\label{sec:pop_relax}
}
Since constraints \eqref{eq:csn_io} and \eqref{eq:cns_param} are bilinear, constraints \eqref{eq:cns_bnd} are linear, and, as stated in Section II, the functional $\mathcal{J}$ is polynomial, problem \eqref{min_tustin_miso} is a POP. Thus, the application of the results presented by \cite{lass04} allows us to build a hierarchy of semidefinite programming (SDP) relaxations \revone{of increasing size, indexed by an integer $\rho \in \mathbb{N}_+$ called \textit{order of relaxation}. The sequence of optimal solutions of the relaxed problems} is guaranteed to converge to the global minimum of \eqref{min_tustin_miso}, \norev{as detailed in the following.}
Unfortunately, the original approach proposed by \cite{lass04}, \revone{yields SDPs with $O((n_un_yN)^{2 \rho})$ optimization variables, which is} computationally cumbersome even if the number $N$ of collected data is relatively small. 

\revone{In this section, we demonstrate that Problem \eqref{min_tustin_miso} enjoys peculiar sparsity properties that allows us to drastically reduce the} computational complexity of the SDP relaxed problems, namely \textit{correlative sparsity}, discussed in \cite{waki06}, and \textit{term sparsity}, analyzed in \cite{WML21a,WML21b}. 

\revone{\begin{result}Let $\mathcal{J}=\pm \theta_i$ for some $i\in \{1,\dots,n_\theta\}$ and $\psi \in \R^{n_\psi}$ be the collection of variables $\gamma,\xi$. There exists $J \in \mathbb{N}_+$ and index sets $\mathcal{I}_r$ and $\mathcal{S}_r$ with $r = 1,\dots,J$ satisfying the running intersection property, i.e.,:\\
\textbf{CS1.}~The set of variables indices $\mathcal{I}_0 \doteq \{1,2,\dots,n_\theta+n_\psi+n_u\,(n_y+1)\,N\}$ is the union of the sets $\mathcal{I}_r$.\\
\textbf{CS2.}~The set of constraints indices $\mathcal{S}_0 \doteq \{1,2,\dots,N_s\}$ (with $N_s\doteq N(n_u+n_y)+ \sum_{m=1}^{n_y}\sum_{l=1}^{n_u} N+n_{m,l}$) is the union of the sets $\mathcal{S}_r$.\\
\textbf{CS3.}~The sets $\mathcal{S}_r$ are mutually disjoint.\\
\textbf{CS4.}~For every $s \in \mathcal{S}_r$, the polynomial constraint $h_s$ depends only on the variables $x(\mathcal{I}_r) = \{x_i : i \in \mathcal{I}_r \}$.\\
\textbf{CS5.}~The objective function $\mathcal{J}$ of Problem \eqref{min_tustin_miso} depends only on the variables $x(\mathcal{I}_r) = \{x_i : i \in \mathcal{I}_r \}$.\\
\textbf{CS6.}~For every $r=1,\dots,J-1:$
    \begin{equation}
        \mathcal{I}_{r+1} \cap \bigcup_{j=1}^r \mathcal{I}_{j} \subseteq \mathcal{I}_{r}.
    \end{equation}
\end{result}
\begin{proof}
    We proceed along the same arguments exploited in \cite{cer12,cer18}. Let the optimization variables be collected (with order) in the vector $x = [$ $\theta_1,\dots,\theta_{n_\theta}$, $\psi_1,\dots,\psi_{n_{\psi}}$, $z_{11}(1),\dots,z_{11}(N)$, \dots, $z_{n_yn_u}(1),\dots,z_{n_yn_u}(N)$, $u_{1}(1),$ $\dots,u_{1}(N)$, $u_{n_u}(1),\dots,$ $u_{n_u}(N)$ $]$. Let $\overline{n}=\max_{m,l} \{n_{m,l}\}$. Take index sets 
    \begin{subequations} \begin{align}
        I_1 &\doteq \{1,\dots,n_\theta+n_\psi\}\\
        I_{2+k} &\doteq \{n_\theta+1,\dots,n_\theta+n_\psi\} \cup V_k 
    \end{align}\end{subequations}
    for $k=1,\dots,N-\overline{n}$, where, given $n_b = n_u (n_y+1)-1$,
    \begin{equation}
        V_k \doteq \bigcup_{i=0,\dots,n_b} \{n_\theta+n_\psi+iN+k,\dots, n_\theta+n_\psi+iN+k+\overline{n}\}.
    \end{equation}
   Next, we notice that
   \begin{itemize}
       \item For all $1\leq m\leq n_y$ and $1\leq l\leq n_u$, constraints \eqref{eq:csn_io} (depending on index $k=n_{m,l},\dots,N$), are only functions of the variables $x(\mathcal{I}_{2})$ for $k=n_{m,l},\dots,\overline{n}$ and of the variables $x(\mathcal{I}_{2+k-\overline{n}})$ for $k=\overline{n}+1,\dots,N$.
       \item For all $1\leq m\leq n_y$, $1\leq l\leq n_u$, and $j=0,\dots,n_{m,l}-1$, constraints \eqref{eq:cns_param} are only functions of the variables $x(\mathcal{I}_1)$.
       \item Constraints \eqref{eq:cns_bnd} (depending on index $k=1,\dots,N$), are only functions of the variables $x(\mathcal{I}_{2})$ for $k=1,\dots,\overline{n}$ and of the variables $x(\mathcal{I}_{2+k-\overline{n}})$ for $k=\overline{n}+1,\dots,N$, because all entries of the vectors $z_{:,l}$ and $u$ indices are contained in the same index set.
       \item The objective function only depends on one optimization variable.
   \end{itemize}
   Building constraints index sets accordingly, the result follows.
\end{proof}

\revthree{By exploiting the correlative sparsity through the use of variable indices $I_r$, we can formulate a reduced-size moment hierarchy of SDPs as
\begin{subequations}\label{eq:cssos}\begin{align}
    \min_{y}~~ & L_y(\mathcal{J}) \\
     \text{\rm s.t.}~~  & y_0=1,\\
     & M_\rho(y,\mathcal{I}_r) \succeq 0, ~~~ r=1,\dots,J\\
     & M_{\rho-d_s}(h_s y,\mathcal{I}_r) \succeq 0, ~ s\in \mathcal{S}_r,~ r=1,\dots,J,
\end{align}\end{subequations}
where $h_s,d_s$ represent the $s$-th constraint of problem \eqref{min_tustin_miso} and its polynomial degree respectively, $y$ are variables representing the entries of standard monomial basis for polynomials up to order $2\rho$, $L_y$ is a linear function, $M_\rho(y,\mathcal{I}_r)$ is the moment submatrix associated with variables $x(\mathcal{I}_r)$, and $M_\rho(h_s y,\mathcal{I}_r)$ is the localizing moment submatrix associated with variables $x(\mathcal{I}_r)$ and constraint $h_s$. We refer the reader to \cite{Las_book09} for a detailed definition of moment matrices.} Specifically, the size of each constraint in \eqref{eq:cssos} is reduced from $O((n_un_y N)^{2\rho})$ to $O((n_un_y \overline{n})^{2\rho})$.

Next, we show that a further reduction in computational complexity is enabled by the following property result:\\
\begin{result}
    Both the objective function $\mathcal{J}$ and the constraints $h$ of Problem \eqref{min_tustin_miso} enjoy term sparsity, i.e., the polynomials $\mathcal{J}$ and $h_s$, $s=1,\dots,N_s$, are formed by a limited number of monomials compared to the full monomial basis for the polynomials in $x(\mathcal{I}_r)$ (with $s \in \mathcal{S}_r$) of the corresponding degree.  
\end{result}
\begin{proof}
    The result immediately follows from inspection of Problem \eqref{min_tustin_miso}.
\end{proof}
     

As described in \cite{lass20}, we can combine techniques based on term sparsity (see, e.g., \cite{WML21b}) with those based on correlative sparsity (see, e.g., \cite{waki06}) to further reduce the size of the SDP \eqref{eq:cssos}. \revthree{In particular, given an order of relaxation $\rho$ and an index set $\mathcal{I}_r$, by suitably defining a sequences of binary matrices $\{B_{\rho,r,0}^{(t)}\}_{t \geq 1}$ and $\{B_{\rho,r,s}^{(t)}\}_{t \geq 1}$ for each constraint $h_s$, we have
\begin{subequations}\label{eq:cs-tssos}\begin{align}
    \min_{y}~~ & L_y(\mathcal{J}) \\
     \text{\rm s.t.}~~  & y_0=1,\\
     & B_{\rho,r,0}^{(t)} \circ M_\rho(y,\mathcal{I}_r) \succeq 0, ~~ r=1,\dots,J\\ 
     \begin{split}
       B_{\rho,r,s}^{(t)} \circ M_{\rho-d_s}(h_s y,\mathcal{I}_r) \succeq 0 \\
    ~~  s\in \mathcal{S}_r,~ r=1,\dots,J,
     \end{split}
\end{align}\end{subequations}
where 
$\circ$ is the Hadamard product. We refer to \cite{lass20} for the exact definition of matrices $B_{\rho,r,s}^{(t)}$.\\
Problem \eqref{eq:cs-tssos} defines a two-level hierarchy of SDP relaxations for Problem \eqref{min_tustin_miso}, indexed by $\rho$ and $t$. This is known as the CS-TSSOS hierarchy. }
The sequence of solutions $\theta^*_{t,\rho}$ generated by the CS-TSSOS hierarchy 
satisfies
\begin{subequations}\begin{align}
    \theta^*_{t,\rho} \leq \theta^*_{t+1,\rho} ~~ t \geq 1, \forall \rho\\
    \theta^*_{t,\rho} \leq \theta^*_{t,\rho+1} ~~ \rho \geq 1, \forall t
\end{align}\end{subequations}
i.e., SDP relaxations provide lower bounds on the global optimum, whose quality improves when increasing $\rho$ and $t$. 
\begin{remark}
    This is not critical for solving Problem \eqref{min_tustin_miso}; in fact, when the goal is to compute the PUIs, the relaxed solutions yield conservative results that are guaranteed to contain the true parameter values.
\end{remark}
Moreover, as shown in \cite[Section 4]{lass20}, under mild assumptions and for any relaxation order $\rho \geq 1$, the sequence of optimal values of Problem \eqref{eq:cs-tssos} generated for $t \geq 1$ converges in finitely many steps to the optimal value of the corresponding problem built using correlative sparsity only, i.e., Problem \eqref{eq:cssos}. In turn, as $\rho \rightarrow \infty$, the sequence of optimal values of \eqref{eq:cssos} converges to the global optimum of \eqref{min_tustin_miso}.
}
\begin{remark}\label{rmk:relax_pui}
Although the convergence of the original SDP relaxation proposed in \cite{lass04} is obtained, in general, for $\rho \rightarrow \infty$, more recent contributions by \cite{mar09} and \cite{nie1} prove that a finite-time convergence is obtained, provided that the problem satisfies a set of mild conditions. We refer the reader to  \cite{nie1} and references therein for details.
\end{remark}

\revtwo{
\subsection{Asymptotic Properties Analysis}
This section investigates the behaviour of the proposed approach as the number of data points increases and its asymptotic properties.

Let $\mathcal{F}_N$ denote the feasible set of Problem \eqref{min_tustin_miso} constructed using $N$ data samples, and let $\Pi_\theta$ denote the projection of a set into the subspace of parameters, i.e., for a set $\mathcal{X} = \{(\theta,\psi,z,u): h(\theta,\psi,z,u) \leq 0\}$, we have $\Pi_\theta(\mathcal{X}) = \{\theta: \exists \psi,z,u: h(\theta,\psi,z,u) \leq 0\}$. The following results hold.} 

\revtwo{
\begin{theorem}[Deterministic inclusion]\label{th:true_in_fps}
    Assume the a-priori information on the system and the bounds on the noise and discretization error are correct. Then, $\mathcal{F}_N$ is guaranteed to contain the true parameter $\theta^\star$ to be estimated, i.e., $\theta^\star \in \mathcal{F}_N$ for all $N \geq n_\theta$. 
\end{theorem}
\begin{proof}
    Under the considered assumptions, $\{ \theta^\star\} \subseteq \mathcal{D}_{\theta,N} \subseteq \Pi_\theta(\mathcal{F}_N)$, where $\mathcal{D}_{\theta,N}$ is the set defined according to Equation~\eqref{eq:fps_generic} using $N$ data points. The latter inclusion is a direct consequence of the observation in Remark~\ref{rmk:subset_F_D}. 
\end{proof}
}

\revtwo{
The significance of this result is that, under the assumption of correct a-priori information on the system and correct bounds, the PUIs are guaranteed to contain the true parameters even if the number of data points is small. This is a substantial advantage compared to the statistical setting, where most results only enjoy an asymptotic validity. Conversely, when some a-priori information is incorrect, data may eventually \textit{falsify} the taken assumptions; i.e., as data increases, $\mathcal{D}$ may become the empty set, reflecting the need for less stringent assumptions in the SM-EIV identification.
}

\revtwo{
Moreover, as the number of available data increases, we can state the following result:
\begin{theorem}[Set-consistency]\label{th:set_consistency}
    Let the assumptions of Theorem~\ref{th:true_in_fps} hold. The sequence of sets $\{\mathcal{D}_N\}_{N \in \mathbb{N}}$ is such that
    \begin{equation}\label{eq:fps_seq_inclusion}
        \{\theta^\star\} \subseteq \Pi_\theta(\mathcal{D}_{N+1}) \subseteq \Pi_\theta(\mathcal{D}_{N}) \dots \subseteq \Pi_\theta(\mathcal{D}_{n_\theta}), ~~ k \geq n_\theta
    \end{equation}
    Moreover, as $N\rightarrow \infty$, the sequence converges to a non-empty limit set $\mathcal{D}_\infty$ such that $\{\theta^\star\} \subseteq \Pi_\theta(\mathcal{D}_\infty)$. 
\end{theorem}
\begin{proof}
    Equation~\eqref{eq:fps_seq_inclusion} follows after noting that, by construction, more constraints are added to the definition of the set when augmenting the data samples, thus reducing its size. The first inclusion is a consequence of Theorem~\ref{th:true_in_fps}. Convergence is ensured thanks to the monotonicity and boundedness of the sequence, and using Equation~\eqref{eq:fps_seq_inclusion}. 
\end{proof}

The significance of this theorem is that, as long as the bounds are correct, increasing the amount of data will lead to an improvement in the accuracy of the estimate. This is captured by the following corollary.

\begin{corollary}[Size-consistency]\label{corollary1}
    Under the assumption of Theorem 1, the size of the PUIs, computed as $\sigma_i \doteq \overline{\theta}_i-\underline{\theta}_i$, is non-increasing with the number of data and converges to a unique limit value $\sigma_i^\infty =  \lim_{N\rightarrow \infty} \sigma_i(N)$.
\end{corollary}
\begin{proof}
    Follows from the application of Theorem 1 and the definition of PUIs.
\end{proof}
}

\revone{Corollary \ref{corollary1} does not ensure point-wise consistency of the estimate, i.e., $\sigma_i^\infty$ is not guaranteed to be zero. Indeed, due to the presence of the discretization error and the measurement noise, such a limit takes a finite non-zero value depending on the bounds $\Delta_\delta(k),\Delta_\eta(k),\Delta_\epsilon(k)$ and the specific dataset.}

\subsection{Comparison with the model transformation approach}
In this section, we discuss the main advantages of the proposed method compared to the model transformation approach previously proposed by \cite{cdc22}.

\norev{We observe that the algorithm in \cite{cdc22} is limited to SISO identification. Consequently, we compare the two algorithms considering SISO models. In this scenario,} the approach proposed in this work requires the solution to a POP with $n^{var}_1 = 4n + 1 + 2N$ optimization variables. At the same time, the POP formulated by \cite{cdc22} involves $n^{var}_2 = 4n + 2(n+1)N$ variables. Therefore, $n^{var}_2$ is significantly larger than $n^{var}_1$ for medium to large values of $n$, hindering the applicability of the method proposed by \cite{cdc22} in those applications where the number of parameters to be estimated is not small. \norev{This is quantitatively demonstrated through the numerical example in Section \ref{sec:simul_order2}. Moreover, we highlight that the model transformation method is practically infeasible for the examples in Sections \ref{sec:simul_order3} to \ref{sec:experim}, due to the significant computational burden.}\\
\norev{Finally, the performance of the algorithm proposed by \cite{cdc22} depends on the user-defined value of the hyperparameter $\tau$.} In contrast, the approach presented in this work does not require any hyperparameter tuning.

\section{Data-driven discretization error bound computation}
\label{sec:discr_estim}
As mentioned in Section~\ref{sec:tustin}, the proposed approach requires knowledge of a bound on the discretization error introduced by the Tustin discretization. In this section, we propose a strategy to estimate this bound by leveraging experimentally collected data and available prior information on model structure and noise bound.

\revone{Whenever we estimate a CT system using a digital procedure and the intersample behavior is unknown, we introduce a discretization error.} Indeed, this is true in classical, probabilistic approaches as well. For a thorough discussion about digital implementation issues in direct CT identification, we refer the reader to \cite{gar03}. 
Due to such an error, CT identification methods cannot accurately estimate the true parameters even when no measurement noise affects the data, as demonstrated in the examples proposed in the CONTSID toolbox demo by \cite{gar18}. 

As noticed by \cite{cdc22}, neglecting the discretization error can lead to the formulation of infeasible POPs. 
\norev{This observation suggests that a reasonable data-driven procedure to estimate the bounds $\Delta_\delta(k)$ is to find the solution to a particular feasibility problem in which we look for the smallest $d$ such that the feasible set of the POP \eqref{min_tustin_miso} is not empty.} The following results hold. 

\begin{result}[Discretization error bound estimation for MISO systems] 
The minimum value of the discretization error \revone{bounds $\Delta_\delta(k)$} such that problem \eqref{min_tustin_miso} is feasible is the solution to the following optimization problem:
\allowdisplaybreaks 
\begin{subequations}   \label{eq:feas_miso} \begin{align}
    &\Delta_\delta = \min_{\alpha,\beta,\gamma,\xi,u,\tau}d \\
    & \quad \text{\rm subject to:} \notag \\
    & \quad z_{m,l}(k) + \sum_{i=1}^{n_{m,l}} \gamma_i^{(m,l)} z_{m,l}(k+i-n_{m,l}) = \\ &\quad =\sum_{i=1}^{n_{m,l}} \xi_i^{(m,l)} u_l(k+i-n_{m,l}), \quad m=1,\dots,n_y, \notag \\
    & \qquad l=1,\dots,n_u,\quad k=n_{m,l}+1,\dots,N \notag \\
    & \quad \left.\begin{aligned}
    & \xi_j^{(m,l)} [a_n(\alpha^{(m,l)})] = b_j(\beta^{(m,l)}) \\
    & \gamma_j^{(m,l)} [a_n(\alpha^{(m,l)})] = a_j(\alpha^{(m,l)})  
     \end{aligned} \right\} ~ \begin{aligned} {m }&{= 1,\dots,n_y,} \\ l&=1,\dots,n_u, \\ j &= 0,\dots,n_{m,l}-1. \end{aligned} \\
    & \quad \underline{\Delta_\eta}(k)-\revone{d\,\tilde \phi_k} \leq \tilde{y}(k) - \sum_{l=1}^{n_u}z_{:,l}(k) \leq \overline{\Delta_\eta}(k) +\revone{d\,\tilde \phi_k} \notag \\
    & \quad \underline{\Delta_\epsilon}(k) \leq \tilde{u}(k) - u(k) \leq \overline{\Delta_\epsilon}(k) \\
    & \qquad k = 1,\dots,N \notag
\end{align} \end{subequations}
\end{result}

\begin{remark}
  \norev{When we solve \eqref{eq:feas_miso}, we look for an upper bound of the value $d^*$ that guarantees feasibility. For this motivation, unlike what was stated in Remark~\ref{rmk:relax_pui}, SDP relaxation methods are unsuitable to solve this problem unless convergence is achieved. Conversely, we can obtain the desired upper bound by any local optimization algorithm that finds a suboptimal feasible solution.} 
\end{remark}

\revone{
\begin{remark}
    We emphasize that our approach does not rely on measurements of the signal's derivatives, nor does it assume any prior knowledge of the system's intersample behavior, apart from signals' boundedness.
\end{remark}
\begin{remark}
    The bounds $\Delta_\delta(k) = d^*\,\tilde \phi_k$ computed using the optimal value $d^*$ of Problem \eqref{eq:feas_miso} are not guaranteed to overestimate the actual discretization error. In fact, since ${\Delta_\delta}(k)$ enters Problem~\eqref{min_tustin_miso} added to $\overline{\Delta_\eta}(k), \underline{\Delta_\eta}(k)$, when $\overline{\Delta_\eta}(k), \underline{\Delta_\eta}(k)$ are not tight descriptions of the actual noise realization, $d^*$ is underestimated by an amount depending on the degree of conservativeness of the bounds $\overline{\Delta_\eta}(k), \underline{\Delta_\eta}(k)$. However, this is not critical since the estimation performed through Problem~\eqref{min_tustin_miso} only depends on the sums $\underline{\Delta_\eta(k)}-\Delta_\delta(k)$ and $\overline{\Delta_\eta(k)}+\Delta_\delta(k)$.
\end{remark} 
}
\section{Numerical results}\label{sec:numerical}
This section proposes three numerical examples with simulated and experimental data. All the computations are performed on a PC with AMD Ryzen 7 1700 Eight-Core Processor @$\SI{3.00}{\giga\hertz}$, $\SI{16}{\giga\byte}$ DDR4 RAM, and running Windows 10.\\
\revthree{For all examples, we solve \eqref{eq:feas_miso} through the local solver Ipopt by \cite{Ipopt}. Next, we employ the softwares TSSOS by \cite{lass20} with order of relaxation $\rho=2$, and Mosek \cite{mosek} to respectively build the SDP relaxation of \eqref{min_tustin_miso} and to compute its solution.}

\subsection{Identification of a second-order SISO system}
\label{sec:simul_order2}
In the first experiment, we consider the system $$ H(s) = \frac{10.5 s - 21}{s^2 + 2.2 s + 16.3},$$ to compare the proposed approach with the method presented by \cite{cdc22}.  The input signal is set to
$$u(t) = \sum_{k=0}^{K} A_k \cos{(k \Delta \omega t + \phi_k)} $$
where $A_k$ and $\phi_k$ are zero-mean normally distributed with unitary variance, $\Delta \omega = \SI{0.1}{\radian\per\second}$ and $K=50$.
We collect $N = 80$ samples of $\Tilde{u}(k) = u(k)$ and $\Tilde{y}(k) = y(k) + \eta(k)$ using sampling period $T_s = \SI{0.05}{\second}$. The output noise is selected such that $|\eta(k)| \leq 2$ for all $k=1,\dots,N$.

\begin{table}[h]
 \centering 
 \caption{Example 1: PUIs and central estimate provided by the proposed method.}
  \begin{tabular}{c|c c c c } 
  & $ \theta_i^{\rm true} $ & $ \underline{\theta_i} $ & $ \theta_i^{\rm C} $ & $ \overline{\theta_i} $\\ 
 \hline 
  $ \alpha_0 $  & $16.3$   & $15.984$   & $16.313$   & $16.643$   \\ 
  $ \alpha_1 $  & $2.2$   & $2.096$   & $2.187$   & $2.278$   \\ 
  $ \beta_0 $  & $-21.0$   & $-21.435$   & $-20.726$   & $-20.017$   \\ 
  $ \beta_1 $  & $10.5$   & $9.915$   & $10.442$   & $10.969$   \\ 
  \end{tabular} 
\label{tab:siso_order2_tustin}
 \end{table} 
 
Applying the proposed approach, we obtain PUIs in Table~\ref{tab:siso_order2_tustin}. 
For comparison, we also estimate the parameter bounds through the model transformation method proposed by \cite{cdc22}. 
Interestingly, all bounds provided by the two methods are the same up to a maximum absolute difference of $7\times 10^{-3}$. This is explained by noting that both approaches provide an approximate reformulation of the set \eqref{eq:fps_generic}. However, the Tustin discretization method drastically reduces computational cost compared to the model transformation approach, with a total runtime of \SI{29.06}{\second} versus \SI{1031.98}{\second}, and memory usage reduced by a factor of approximately 19.5. This highlights the superior efficiency of the Tustin method.

\subsection{Identification of a third-order SISO system}
\label{sec:simul_order3}
In the second numerical experiment, we generate noisy numerical data by simulation of the CT system described by the transfer function 
$$ H(s) = \dfrac{11 s^2 + 36 s + 103}{s^3 + 8 s^2 + 20 s + 100}.$$
We use input signal $$u(t) = \sum_{k=0}^{K} A_k \cos{(k \Delta \omega t + \phi_k)} $$ where $A_k$ and $\phi_k$ are zero-mean normally distributed with unitary variance, $\Delta \omega = \SI{0.1}{\radian\per\second}$ and $K=120$. This kind of excitation is widely used in CT identification; see, e.g., \cite{hallemans_frf_2022}. We collect $N = 300$ samples of $\Tilde{u}(k) = u(k T_s) + \epsilon(k)$ and $\Tilde{y}(k) = y(k T_s) + \eta(k)$, where $|\eta(k)| \leq 2.2$ and $|\epsilon(k)| \leq 0.3$, for all $k=1,\dots,N$, which corresponds to input and output SNR equal to $\text{SNR}_u = \SI{32.8}{\decibel}$ and $\text{SNR}_y = \SI{23.3}{\decibel}$, respectively. The selected sampling period is $T_s = \SI{0.03}{\second}$.

Following the proposed approach, we estimate $\Delta_\delta(k) = d^* = 0$ as an absolute bound on the discretization error, indicating that the bound on the measurement noise is sufficient to capture both measurement noise and discretization error for the given dataset. Then, using this value, we compute the PUIs. We compare the central estimate $\theta^C$ with the result obtained by using the state-of-the-art CT identification toolbox CONTSID \cite{gar18}.

\begin{table}[ht] 
 \centering 
 \caption{Example 2: PUIs, central estimate, and comparison with the SRIVC algorithm.}
  \begin{tabular}{c | c c c c | c } 
  & $ \theta_i^{\rm true} $ & $ \underline{\theta_i} $ & $ \theta_i^{\rm C} $ & $ \overline{\theta_i} $ &  $ \theta_i^{\rm SRIVC} $ \\ 
 \hline 
  $ \alpha_0 $  & $100$   & $76.753$   & $106.947$   & $137.141$ & $111.329$ \\ 
  $ \alpha_1 $  & $20$   & $18.522$   & $20.508$   & $22.495$ & $21.188$  \\  
  $ \alpha_2 $  & $8$   & $6.447$   & $8.460$   & $10.473$  &  $8.742$ \\ 
  $ \beta_0 $  & $103$   & $82.030$   & $109.054$   & $136.077$  & $121.209$  \\ 
  $ \beta_1 $  & $36$   & $27.271$   & $39.490$   & $51.708$  &$ 44.567$  \\ 
  $ \beta_2 $  & $11$   & $10.231$   & $11.299$   & $12.367$  & $11.811$ 
  \end{tabular} 
\label{tab:siso_order3_tustin}
 \end{table} 

Table~\ref{tab:siso_order3_tustin} reports the computed PUIs and compares the central estimates obtained through the Tustin discretization method and the best CONTSID estimate provided by the SRIVC algorithm by \cite{GARNIER2007}. 
Notice that each PUI contains the corresponding true parameter, a key feature of SM methods. Moreover, we measure the accuracy in terms of the average relative error, defined as $ \frac{1}{6} \sum_{i=1}^6 \text{rel.err}_i$, where $\text{rel.err}_i = 100\frac{|\theta_i^{\rm true}-\hat{\theta}_i|}{|\theta_i^{\rm true}|}$ and $\hat{\theta}$ represents the parameter estimate. The average relative error of the proposed approach is $5.58\%$, while that of $\theta^\text{SRIVC}$ is $12.56\%$, indicating that the Tustin discretization method achieved a better accuracy compared to SRIVC. The improved accuracy comes at the cost of an increased computational effort, which is the main drawback of the proposed approach. More precisely, the proposed method requires \SI{4387.25}{\second}, while SRIVC only requires \SI{2.4}{\second}.

\subsection{Identification of a seventh-order SISO system}
\label{sec:simul_order7}
In this section, we consider a SISO system with seven poles and no zeros to test a higher-order example with high relative degree. The system's poles are located at $-1 \pm 0.6325i$, $-1.1$, $-1$, $-0.95$, $-0.9$ and $-0.8$, while its DC-gain equals $9.493$. 

We generate the data by simulation using the input signal $$u(t) = \sum_{k=0}^{K} A_k e^{-b_k \vert t-k \vert^2}$$, where $A_k$ is zero-mean normally distributed with unitary variance and $b_k$ is uniformly distributed in $[0,7]$. The input and output signals are sampled at a period of $T_s = \SI{0.15}{\second}$, resulting in $N = 227$ data pairs. 
The output data are corrupted with additive noise bounded according to $\vert \eta(k) \vert \leq \Delta_\eta(k) = 0.05 \vert y(k) \vert$, while input data are not corrupted by noise (i.e., we consider the output error noise setting). The output signal-to-noise ratio is \SI{30.8}{\decibel}.

We estimate the bound on the \textit{relative output} discretization error as $\Delta_\delta(k) = 0.0012 \vert \Tilde{y}(k) \vert$ using Problem \eqref{eq:feas_miso}. Next, we perform the identification exploiting the prior knowledge about the system's relative degree. Accordingly, we compute the PUIs solving Problem \eqref{min_tustin_miso} after setting $\beta_1=\beta_2=\dots=\beta_7=0$ and using $\mathcal{J} = \theta_i$ and $\mathcal{J} = -\theta_i$ for $\theta_i \in \{\beta_0,\alpha_0,\dots,\alpha_7\}$. This formulation enhances the accuracy of the estimated model as the additional a-priori information about the numerator parameters allows reducing the feasible set size.

\begin{table}[h!]
\centering
\caption{Example 3: PUIs, central estimate, and comparison with SRIVC algorithm.}
\label{tab:es_highorder}
\begin{tabular}{c|c|ccc|c}
& True & $\underline{\theta}_i$ & ${\theta}_i^c$ & $\underline{\theta}_i$ & $\theta_i^{\text{SRIVC}}$ \\
\hline
$\beta_0$ & 10 & 6.73 & 10.22 & 13.71 & 3.35 \\
$\alpha_0$ & 1.05  & 0.71 & 1.07  & 1.44  & 0.36 \\
$\alpha_1$ & 7.11  & 4.79 & 7.27  & 9.74  & 2.44 \\
$\alpha_2$ & 20.67 & 13.91 & 21.17 & 28.42 & 8.49 \\
$\alpha_3$ & 33.61 & 22.83 & 34.31 & 45.78 & 13.29 \\
$\alpha_4$ & 33.15 & 22.12 & 34.07 & 46.02 & 18.07 \\
$\alpha_5$ & 19.9 & 14.05 & 20.30 & 26.55 & 10.43 \\
$\alpha_6$ & 6.75  & 4.35 & 7.11  & 9.88  & 6.89 \\
\end{tabular}
\end{table}

Table \ref{tab:es_highorder} compares the obtained results with those provided by the SRIVC method implemented in the CONTSID toolbox by \cite{gar18}. We notice that the proposed approach provides correct bounds and a quite accurate central estimate. Conversely, due to the limited amount of data, the SRIVC estimate $\theta_i^{\text{SRIVC}}$ is significantly less accurate, thereby demonstrating, as expected, the superior data efficiency of the proposed SM-EIV method.

\subsection{MIMO identification of an electronic circuit}
\label{sec:experim}
\begin{figure}[ht]
    \centering
    \includegraphics[width=\linewidth]{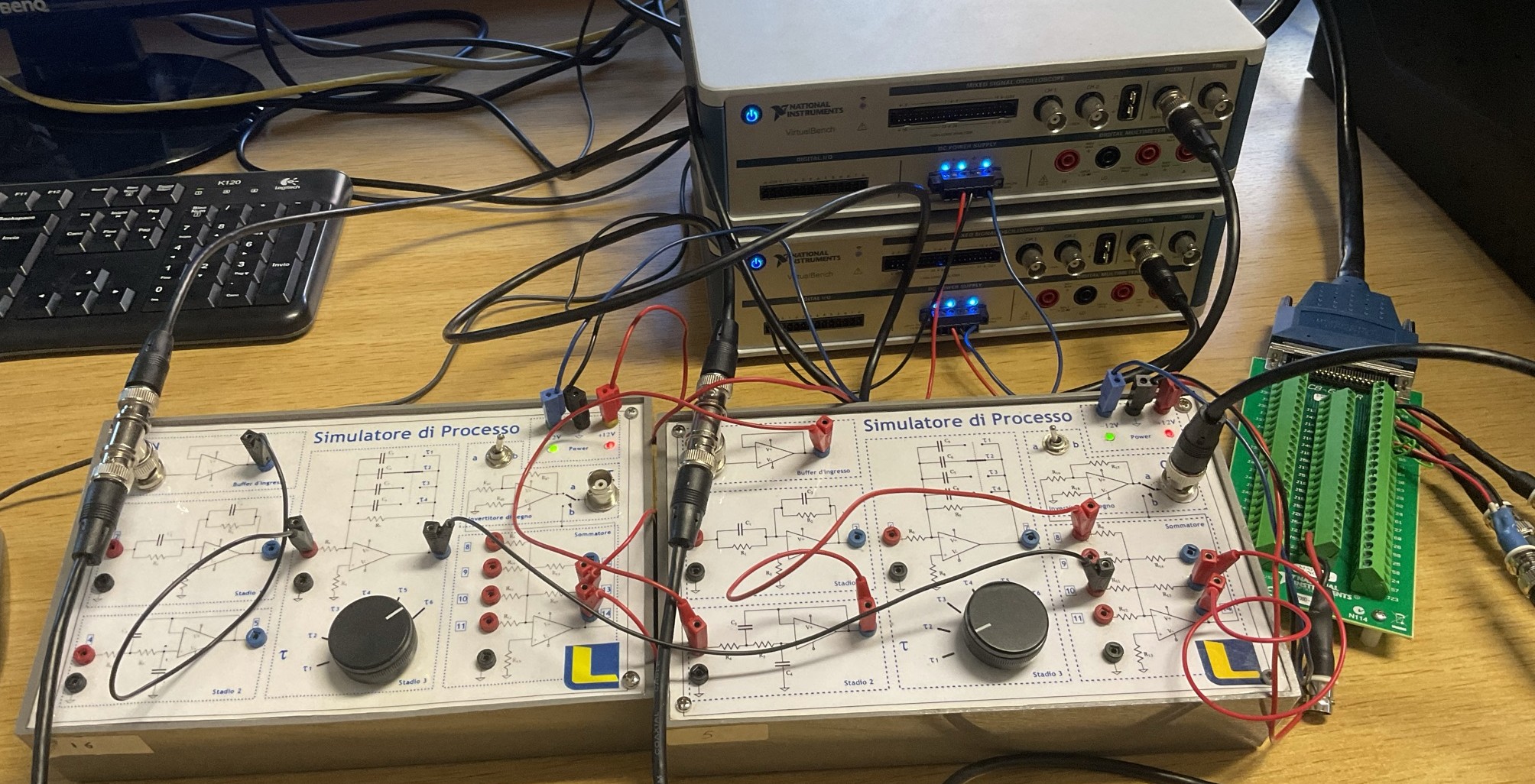}
    \caption{Experimental setup.}
    \label{img:experim}
\end{figure}

In this section, we report the results from an experimental test of the \norev{proposed procedure}. We build the system under study using filters based on operational amplifiers. In particular, we realize a two-input single-output (TISO) circuit by adding the outputs from a first-order low-pass filter and a second-order Sallen-Key low-pass filter through an adder circuit. Figure \ref{img:experim} shows the experimental setup.

From the accuracy of the measurement equipment, we derive the
upper bounds on the measurement errors, $\Delta_{\epsilon,1} = \Delta_{\epsilon,2} = \Delta_{\eta} = 0.015.$
To identify the system, we collect $N = 250$ samples from the input and output sequences through the National-Instruments PXI, equipped with an NI-6221 DAQ board and using a sampling rate $f_s = \num{1e-3}$ {\rm Hz}. Both inputs are \norev{signals of the form $$u(t) = \sum_{k=0}^{K} A_k \cos{(k \Delta \omega t + \phi_k)}, $$ where $A_k,\phi_k$ are normally distributed with unitary variance, $\Delta \omega = \SI{0.1}{\radian\per\second}$ and $K=200$. \\
Applying the procedure described in Section \ref{sec:discr_estim}, we estimate the bound on the discretization error as $\Delta_\delta(k) = d^* \leq 0.05$. Then, we consider two different estimators to identify the system. First, we evaluate the PUIs and the Chebyshev center $\theta^C$ of the FPS. Second, we evaluate the estimate $\theta^{E}$ that minimizes the simulation error within the FPS, that is, in \eqref{eq:sm_generic} we set $\mathcal{L}(\theta) = \sum_{k=1}^N{ \left( \tilde{y}(k) - \hat{y}(k,\theta) \right)^2}$, where $\hat{y}(k,\theta)$ is the output of the identified model.}
%
In Table \ref{tab:experim}, we report (i) the parameters $\theta^{\rm fp}$ computed by using the first principles of physics and the nominal values of the electric components; (ii) the bounds on each parameter $\underline{\theta}, \overline{\theta}$ and central estimate $\theta^C$; (iii) the conditional minimum simulation error $\theta^{E}$.
The time required to compute the PUIs and $\theta^C$  is $\SI{8548.92}{\second}$, while the time to compute $\theta^E$ is $\SI{687.31}{\second}$.
\begin{table*}[ht]
    \centering
    \caption{Identification of the MIMO electronic circuit: estimated parameters}
    \begin{tabular}{c| c | c c c | c }
           & $\theta_i^{\rm fp}$  &  $ \underline{\theta_i} $ & $ \theta_i^{\rm C} $ &  $ \overline{\theta_i} $ & $\theta_i^{\rm E}$  \\
        \hline
        $\alpha^{(1)}_1$  & $13.297$        & $15.372$ & $ 16.186$ & $17.001$                   & $16.292$  \\
        $\alpha^{(1)}_0$  & $4420.8$      & $4559.1$ & $ 4607.9$ & $4656.8$             & $4632.5$   \\
        $\beta^{(1)}_1$   & $0$             & $-0.5467$ & $1.5026$ &   $3.5521$                 & $1.611$  \\
        $\beta^{(1)}_0$   & $4420.8$      & $4601.2$ & $4698.9$ & $4796.7$              & $4689.6$  \\
        \hline
        $\alpha^{(2)}_0$  & $10$            & $10.462$ & $11.311$ &  $12.159$                   & $10.939$   \\
        $\beta^{(2)}_0$   & $-10$           & $-12.643$ & $-10.698$ & $-8.753$                  & $-10.591$  
    \end{tabular}
    \label{tab:experim}
\end{table*}

Finally, we perform two validation experiments by setting the inputs as described in Table \ref{tab:experim_test}. We use the symbols $u^I_1,\ u^I_2$ and $u^{II}_1,\ u^{II}_2$ to indicate the input signals applied in the validation experiments $I$ and $II$, respectively. 
We collect input and output data with sampling period $T_s^{\rm valid} = \SI{2e-5}{\second}$. In Table \ref{tab:experim_test}, we also report the accuracy of the identified model in terms of the following performance indexes:
\begin{align*}
    MSE &= \dfrac{1}{N} \sum_{k=1}^N (\hat{y}(k) - \tilde{y}(k) )^2     \\
    FIT &= 1 - \sqrt{ \dfrac{ \sum_{k=1}^N (\hat{y}(k) - \tilde{y}(k) )^2 }{ \sum_{k=1}^N (\tilde{y}(k) - m_y )^2 } }
\end{align*} 
where $\hat{y}(k)$ is the simulated output and $m_y = \dfrac{1}{N} \sum_{k=1}^N \tilde{y}(k)$.
\begin{table}[ht]
    \centering
    \caption{Identification of a MIMO electronic circuit: results obtained on the validation data sets}
    \begin{tabular}{c|c | c | c }
        Inputs & Model & MSE & FIT  \\
        \hline 
        \rule{0pt}{0.3cm} \multirow{2}{*}{\shortstack{$u^I_1=$ square wave  \\ $u^I_2=$ sawtooth wave}}    & $H(s|\theta^C)$  &  $0.067$ & $94.04\%$ \\
        \cline{2-4}
        \rule{0pt}{0.3cm}    & $H(s|\theta^E)$  &  $0.071$ & $93.89\%$ \\
        \hline
        \rule{0pt}{0.3cm}\multirow{2}{*}{\shortstack{$u^{II}_1=$ sawtooth wave  \\ $u^{II}_2=$ triangular wave}}  & $H(s|\theta^C)$ & $0.016$ & $91.32\%$ \\
        \cline{2-4}
        \rule{0pt}{0.3cm}  &  $H(s|\theta^E)$  &  $0.007$ & $94.24\%$
    \end{tabular}
    \label{tab:experim_test}
\end{table}

\begin{figure*}
    \begin{subfigure}{0.5\textwidth}
    \centering
    \includegraphics[width=\linewidth]{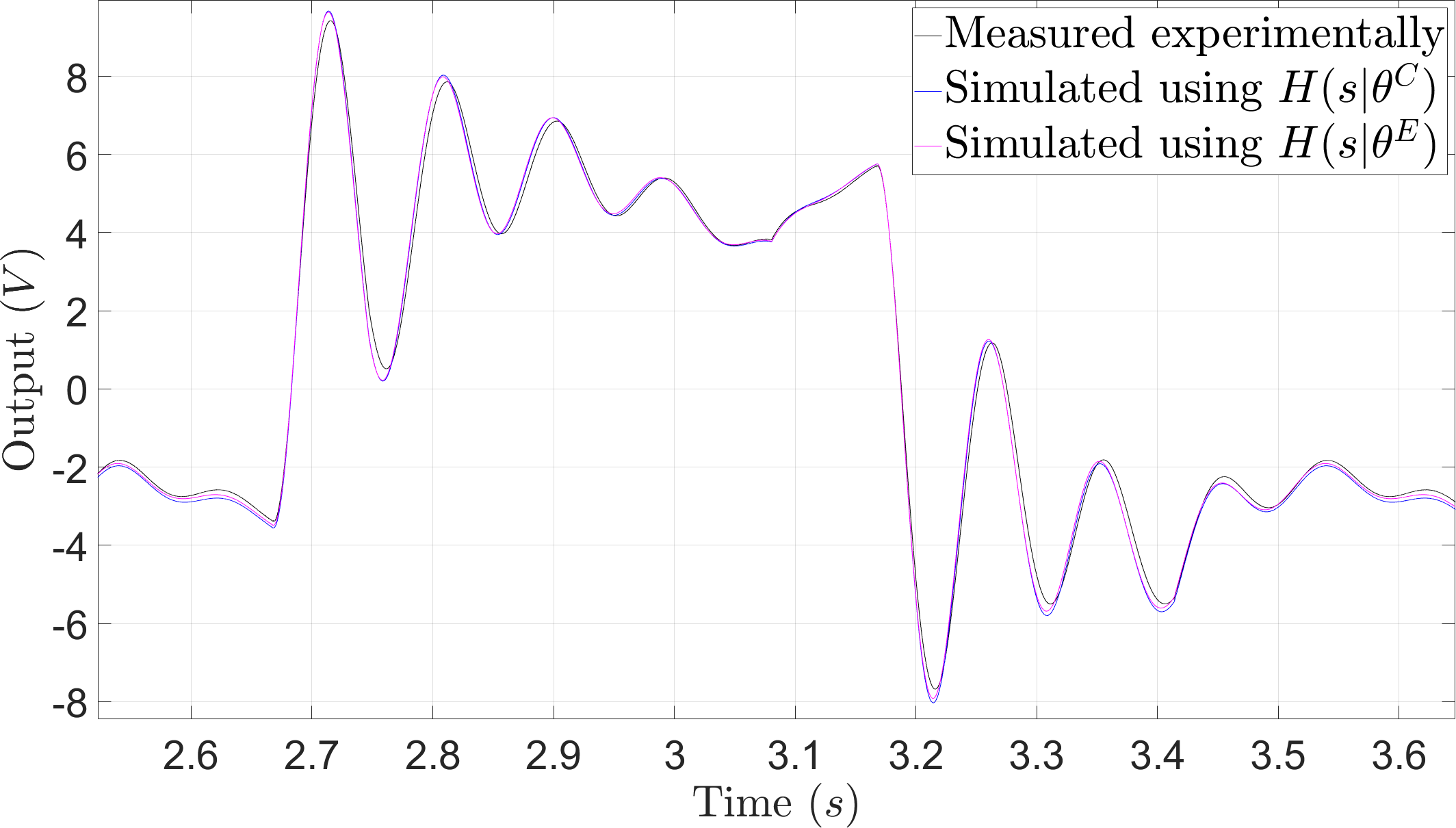}
    \caption{Validation experiment 1}
    \label{fig:test1}
    \end{subfigure}
    \hfill 
    \begin{subfigure}{0.5\textwidth}
    \centering
    \includegraphics[width=\linewidth]{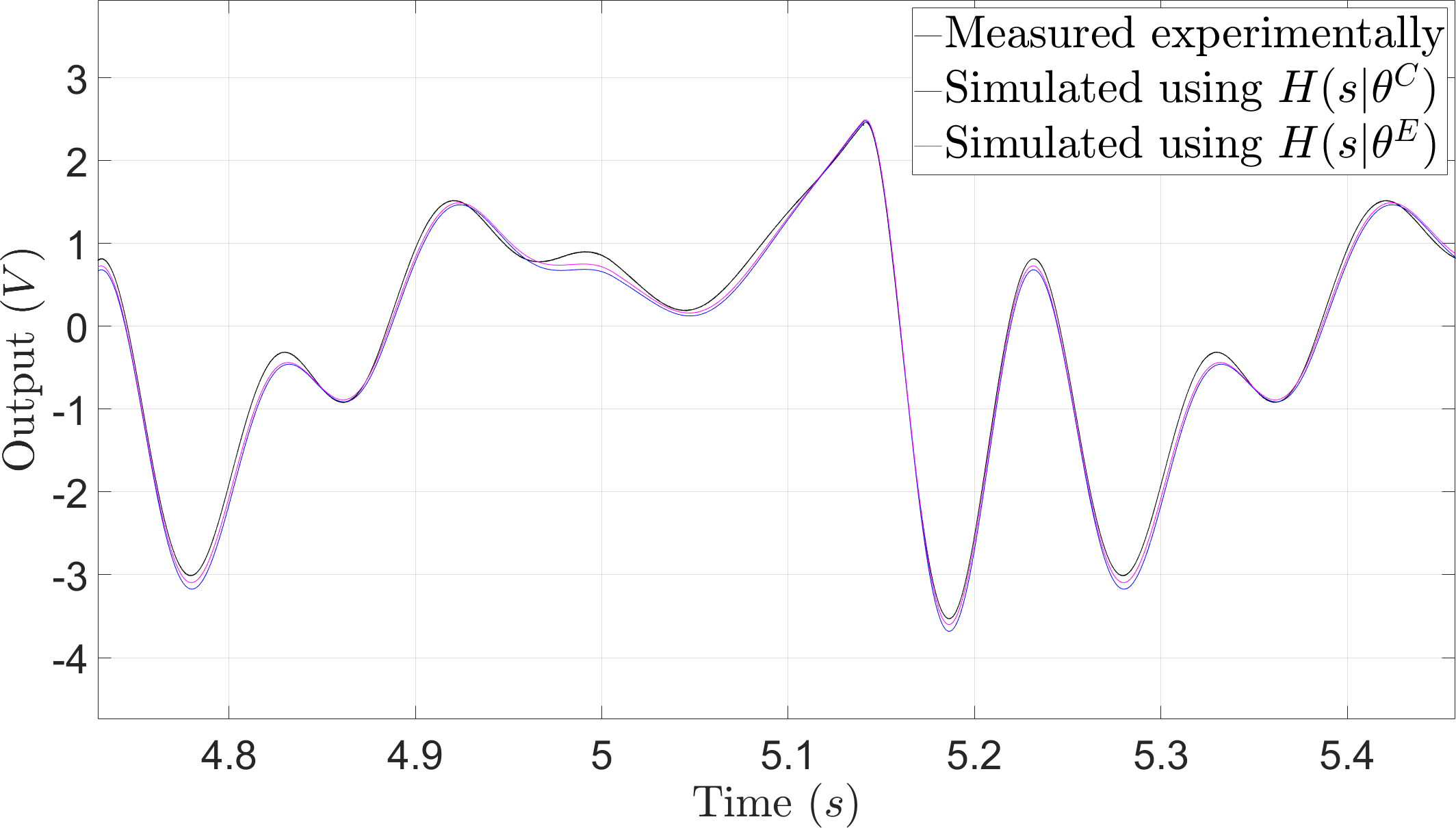}
    \caption{Validation experiment 2}
    \label{fig:test2}
    \end{subfigure}
    \caption{Comparison of measured (black) and simulated (blue $\theta^C$ and red $\theta^E$) output}
    \label{fig:test}
\end{figure*}

Figure~\ref{fig:test} show the comparison between the measured outputs and the simulated outputs over a time interval corresponding to a single generic period of the inputs.
Table \ref{tab:experim_test} and Figure~\ref{fig:test} show that the identified models accurately describe the behaviour of the true system.

\section{Conclusion}
\label{sec:concl}
In this paper, we deal with the problem of estimating the parameters of the matrix transfer function of a MIMO continuous-time LTI system from experimentally collected samples of the input and output signals. By assuming that both the input and output samples are corrupted by bounded noise, we propose a set-membership formulation of the problem based on a suitable application of the Tustin transformation. In the considered framework, the estimate of the CT model parameters turns out to be the solution to a polynomial optimization whose global optimal solution can be approximated applying a computationally tractable convex relaxation. Since the proposed procedure requires the knowledge of a bound on the discretization error introduced by the Tustin transformation, we also propose a data-driven procedure for the computation of such a quantity. We successfully test the performance of the proposed approach in two simulation examples where we compare our method to the state-of-the-art algorithms for CT identification available in the literature. The SM method is also applied to the problem of identifying a CT model of a real electronic circuit from a set of experimentally collected data. The obtained results show that the proposed approach can be used to solve effectively real-world modeling problems.

\section{Acknowledgments}

We thank LADISPE Lab at Politecnico di Torino for providing us with the software and the process simulator used for the data acquisition used in the experimental test.

\bibliographystyle{apalike}
\bibliography{the_bib}
\end{document}